\documentclass[fleqn,usenatbib]{mnras}
\usepackage{hyperref} 
\usepackage{graphicx}
\usepackage[T1]{fontenc} 
\usepackage{ae,aecompl}  

\newcommand\un[1]{{\,\rm #1}}
\newcommand\E[1]{\times10^{#1}}
\newcommand\rs[1]{_\mathrm{#1}}
\newcommand\pd[2]{\frac{\partial{#1}}{\partial{#2}}}

\title[Time-dependent DSA and injection]{
Time-dependent shock acceleration of particles. Effect of the time-dependent injection, 
with application to supernova remnants
}
\author[O. Petruk, B. Kopytko]{O. Petruk$^{1,2}$, B. Kopytko$^{3}$\\
$^1$INAF - Osservatorio Astronomico, Piazza del Parlamento, 1, 90134 Palermo, Italy\\
$^2$Institute for Applied Problems in Mechanics and Mathematics, Naukova 3-b, 79060 Lviv, Ukraine\\
$^3$Institute of Mathematics, Czestochowa University of Technology, Dabrowskiego 69, 42201 Czestochowa, Poland
}

\date{Last updated ...; in original form ...}
\pubyear{2016}

\begin{document}

\label{firstpage}
\pagerange{\pageref{firstpage}--\pageref{lastpage}} 
\maketitle

\begin{abstract}
Three approaches are considered to solve the equation which describes the time-dependent diffusive shock acceleration of test particles at the non-relativistic shocks. At first, the solution of Drury (1983) for the particle distribution function at the shock is generalized to any relation between the acceleration time-scales upstream and downstream and for the time-dependent injection efficiency. Three alternative solutions for the spatial dependence of the distribution function are derived. Then, the two other approaches to solve the time-dependent equation are presented, one of which does not require the Laplace transform. 
At the end, our more general solution is discussed, with a particular attention to the time-dependent injection in supernova remnants. It is shown that, comparing to the case with the dominant upstream acceleration time-scale, the maximum momentum of accelerated particles shifts toward the smaller momenta with increase of the downstream acceleration time-scale. The time-dependent injection affects the shape of the particle spectrum. In particular, i) the power-law index is not solely determined by the shock compression, in contrast to the stationary solution; ii) the larger the injection efficiency during the first decades after the supernova explosion, the harder the particle spectrum around the high-energy cutoff at the later times. This is important, in particular, for interpretation of the radio and gamma-ray observations of supernova remnants, as demonstrated on a number of examples.
\end{abstract}


\section{Introduction}

How does a stellar object become a supernova remnant (SNR) after the supernova event? Some hints come from observations of Supernovae in other galaxies \cite[e.g][]{Weiler-et-al-1986} but -- since they are far away -- they are not quite informative for understanding of how young SNRs obtain their look, how properties of the explosion and ambient medium affect their evolution. Different time and length scales need to be treated in order to model the transfiguration of SN to SNR, that creates difficulties for  numerical simulations. In the last years however a number of studies has been performed in order to understand the involved processes. They adopt either one-dimensional simulations \citep[e.g.][]{Badenes-et-al-2008,Patnaude-et-al-2015} or -- quite recently -- three-dimensional models \citep{Orlando-et-al-2015,Orlando-et-al-2016a}.

In order to relate an SNR model to observations, one has to simulate emission. Radiation of the highly energetic  particles is an important component of a model. The particle spectrum has to be known in order to simulate their emission. The {\em non-stationary} solution of the diffusion-convection equation has to be used in order to describe the distribution function $f(t,x,p)$ of these particles in young SNRs because the acceleration is not in the steady-state regime yet. There are evidences from numerical simulations that the particle spectrum could not be stationary even in the rather old SNRs \citep{Brose-et-al-2016}. 

There is well known approach \citep{Drury-1983,Forman-Drury-1983} to derive the time-dependent solution and expression for the acceleration time. The original formulation has been developed i) for the spatially constant flow velocities $u$ and diffusion coefficients $D$ before and after the shock, ii) for the momentum dependence of the diffusion coefficient of the form $D\propto p^{\alpha}$ with the constant index $\alpha$, iii) for the impulsive or the constant particle injection, iv) for the monoenergetic injection of particles at the shock front and v) for the case when the acceleration time upstream $t_1$ is much larger than that downstream $t_2$. \citet{Toptygin-1980} was the first to consider the time-dependent acceleration and has given a solution for $t_1=t_2$ and the diffusion coefficient independent of the particle momentum $p$. \citet{Drury-1991} has presented a way to generalize his own  solution to include also the spatial dependence of the flow velocity $u(x)$ and the diffusion coefficient $D(x)$. 
\citet{Ostr-Schlick-1996} have found a generalization of the \citet{Toptygin-1980} solution ($t_1=t_2$ and momentum independent $D$) which allows one to consider different $t_1$ and $t_2$. They have also obtained the expression for the acceleration time if there are the free-escape boundaries upstream and downstream of the shock. 
\citet{Tang-Chevalier-2015} have generalized the \citet{Toptygin-1980} solution to the time evolution of the pre-existing seed cosmic rays, i.e. the authors have generalized the treatment to the impulsive (at time $t=0$) injection of particles residing in the half-space before the shock and being distributed with some spectrum $Q\rs{p}(p)$. 
The approach to treat the time-dependent non-linear acceleration is developed by \citet{Blasi-et-al-2007} who have not obtained the solution but made an important progress in derivation of the acceleration time for the case when the particle back-reaction on the flow is important. 

In the present paper, the Drury's test-particle approach is extended to more general situations. Namely, few different representations for $f(t,x,p)$ are obtained; a way to avoid the $t_1\gg t_2$ limitation in deriving the distribution function at the shock $f\rs{o}(t,p)$ is presented; a solution is written in a way to allow for any time variation of the injection efficiency; a possibility for the diffusion coefficient to have other than the power-law dependence on momentum is considered. 

The structure of the paper is as follows. The task and main assumptions are stated in Sect.~\ref{kineq2:kineqbase}. The three different approaches to solve the non-stationary equation are presented in Sections~\ref{kineq2:kineqI}, \ref{kineq2:kineqII} and \ref{kineq2:kineqIII} respectively. Then, in Sect.~\ref{kineq2:discussion}, we demonstrate when and to which extent our generalized solution differs from the original Drury's formulation (Sect.~\ref{kineq2:pmax00}) and discuss implications of the time-dependent injection efficiency on the particle spectrum (Sects.~\ref{kineq2:injteffect}, \ref{kineq2:pinter}). Sect.~\ref{kineq2:conclusions} concludes. Some mathematical identities used in the present paper are listed in the Appendix \ref{kineq2:app1}. 

\section{Kinetic equation and assumptions}
\label{kineq2:kineqbase}

We consider the parallel shock and (without loss of generality) the coordinate axis $x$ to be parallel to the shock normal. The shock front is at $x=0$. The flow moves from $-\infty$ to $+\infty$. 

The one-dimensional equation for the isotropic non-stationary distribution function $f(t,x,p)$ is \citep{Skilling1975a,Jones-1990}:
\begin{equation}
 \pd{f}{t}+u\pd{f}{x}=\pd{}{x}\left[D\pd{f}{x}\right]+\frac{1}{3}\frac{du}{dx}p\pd{f}{p}+Q
 \label{kineq:kineq}
\end{equation}
where $t$ is the time, $x$ the spatial coordinate, $p$ the momentum, $D$ the diffusion coefficient, $Q$ the source (injection) term. The equation is written in the reference frame of the shock front. 
The velocities of the scattering centers are assumed to be much smaller than the flow velocity $u$. 
The injection term $Q$ is considered as a product of terms representing temporal, spatial and momentum dependence
\begin{equation}
 Q=Q\rs{t}(t)Q\rs{p}(p)Q\rs{x}(x).
 \label{kineq2:injtermgen}
\end{equation}
In particular, it is $Q\propto H(t)\delta(p-p\rs{i})\delta(x)$ \citet{Drury-1983,Forman-Drury-1983,Ostr-Schlick-1996} and $Q\propto\delta(t)Q\rs{p}(p)H(-x)$, where $H$ is the Heaviside step function, in \citet{Tang-Chevalier-2015}. 

In the present paper, the injection is assumed to be isotropic and monoenergetic with the initial momentum $p\rs{i}$  \citep[e.g.][]{Blasi-2002}:
\begin{equation}
 Q\rs{p}(p)=\frac{\eta n_1u_1}{4\pi p\rs{i}^2}\delta(p-p\rs{i}),
 \label{kineq:umova2b}
\end{equation}
where the parameter $\eta$ is the injection efficiency; it gives the fraction of particles which are accelerated. The particles are injected at the shock front: $Q\rs{x}=\delta(x)$. 
Different representations of the term $Q\rs{t}(t)$ are considered; for example, it is $Q\rs{t}=1$ for the constant injection and $Q\rs{t}\propto\delta(t)$ for the impulsive injection. 

A number of other assumptions are typically used in order to solve the equation. The distribution function is 
\begin{equation}
 f(0,x,p)=0, \quad \mathrm{for}\quad t=0.
 \label{kineq2:umova1}
\end{equation}
The distribution is continuous at the shock: 
\begin{equation}
 f_1(t,p)=f_2(t,p)\equiv f\rs{o}(t,p)
 \label{kineq:umova3}
\end{equation}
where the index `o' represents values at the front ($x=0$), the index `1' denotes the point right before the shock front ($x=-0$) and the index `2' marks the point right after the shock ($x=+0$). 
The distribution function is uniform downstream of the shock:
\begin{equation}
 f(x)=\mathrm{const},\quad x>0.
 \label{kineq:umova4}
\end{equation}
There is no seed energetic particles far upstream:
\begin{equation}
 f(-\infty)=0,
 \label{kineq:umova5a}
\end{equation}
\begin{equation}
 \left.\pd{f}{x}\right|_{-\infty}=0.
 \label{kineq:umova5b}
\end{equation}

The flow velocity is spatially constant before and behind the shock:
\begin{eqnarray}
 u(x)=u_1, & x<0,
 \label{kineq:umova7a} \\
 u(x)=u_2, & x>0,
 \label{kineq:umova7b}
\end{eqnarray}
where both $u_1$ and $u_2$ are positive and constant, $u_1>u_2$. The ratio $\sigma=u_1/u_2$ is the shock compression factor. Eqs.~(\ref{kineq:umova7a})-(\ref{kineq:umova7b}) are related to the `test-particle' regime when the accelerated particles do not modify the flow structure. In this case, the derivative $du/dx$ is 
\begin{equation}
 \frac{du}{dx}=(u_2-u_1)\delta(x).
 \label{kineq:umova-tp}
\end{equation}
In the present paper, we consider the diffusion coefficients $D_1$ and $D_2$ spatially constant in their domains. 

\section{Approach I. Laplace transform of the original equation}
\label{kineq2:kineqI}

In this section, the approach to a solution \citep{Drury-1983,Forman-Drury-1983} is reviewed and generalized. 
The solution for the distribution function at the shock $f\rs{o}(t,p)$ was derived initially under a limiting assumption that the particle acceleration time in the upstream medium is much larger than the acceleration time downstream. In the present section, we show how a more general expression may be obtained. We describe also three ways to write down expressions for the distribution function $f(t,x,p)$ outside the shock. Our generalization of the \citet{Drury-1983} solution allows for any (integrable) dependence of the injection efficiency on time.

The treatment of the equation (\ref{kineq:kineq}) consists in applying the Laplace transform to the equation that leads to an equation for the Laplace transform $\overline{f}$ of the distribution function $f$: 
\begin{equation}
 s\overline{f}+u\pd{\overline{f}}{x}=\pd{}{x}\left[D\pd{\overline{f}}{x}\right]+\frac{1}{3}\frac{du}{dx}p\pd{\overline{f}}{p}+\overline{Q\rs{t}}(s)Q\rs{p}(p)\delta(x),
 \label{kineq2:kineqg}
\end{equation}
Note that $Q\rs{t}(t)=1$ (continuous steady-state injection after $t=0$) was adopted in the original formulation \citep{Drury-1983}; then $\overline{Q\rs{t}}(s)=1/s$. Hereafter, the over-line marks the Laplace transform. 

The function of interest $f(t,x,p)$ is given by the inverse Laplace transform
\begin{equation}
 f(t,x,p)={\cal L}^{-1}\left\{\overline{f}(s,x,p)\right\}
\label{kineq2:solinvlapldef}
\end{equation}
of the solution $\overline{f}$ of the Eq.~(\ref{kineq2:kineqg}).

\subsection{Function $\overline{f}(s,x,p)$}

Before the shock ($i=1$) and after the shock ($i=2$), the equation (\ref{kineq2:kineqg}) simplifies to 
\begin{equation}
 s \overline{f}+u_i\pd{\overline{f}}{x}=\pd{}{x}\left[D_i\pd{\overline{f}}{x}\right].
 \label{kineq2:eqgx}
\end{equation}
We shall look for the solution in the form
\begin{equation}
 \overline{f}(s,x,p)=\overline{f\rs{o}}(s,p)\exp\left(\frac{u_ix}{D_i(p)}\beta_i(s,p)\right),
 \label{kineq2:solxTP}
\end{equation}
where the diffusion coefficients are uniform $D_i(x)=\mathrm{const}$ and  
$\overline{f\rs{o}}(s,p)$ is the value of the function in the point $x=0$. Substitution Eq.~(\ref{kineq2:eqgx}) with (\ref{kineq2:solxTP}) gives $\beta$ upstream and downstream:
\begin{equation}
 \beta_i(s,p)=\frac{1}{2}\left[1\pm\left(1+\frac{4sD_i(p)}{u_i^2}\right)^{1/2}\right].
 \label{kineq2:betadef}
\end{equation}
where the condition $f(x\rightarrow-\infty)=0$ and thus $\overline{f}(x\rightarrow-\infty)=0$ was used. 
The correspondence of the sign '$+$' in (\ref{kineq2:betadef}) to the upstream and the sign '$-$' to the downstream is clearly demonstrated by the limit $t\rightarrow \infty$. Namely, the stationary solution comes from (\ref{kineq2:solxTP})-(\ref{kineq2:betadef}) with substitution $s=0$:
\begin{equation}
 \mathrm{f}(x,p)={\cal L}^{-1}\left\{\overline{f}\right\}=
 \mathrm{f}\rs{o}(p)\exp\left(\frac{u_1x}{D_1}\right),\quad x<0,
 \label{kineq2:stacsolx-up}
\end{equation}
\begin{equation}
 \mathrm{f}(x,p)={\cal L}^{-1}\left\{\overline{f}\right\}=
 \mathrm{f}\rs{o}(p),\quad x>0.
 \label{kineq2:stacsolx-dw}
\end{equation}

\subsection{Function $f\rs{o}(t,p)$}
\label{kineq2:sectfotp}

In order to find the equation for the function $\overline{f}$ on the shock, i.e. for $\overline{f}(s,0,p)\equiv \overline{f\rs{o}}(s,p)$, the equation (\ref{kineq2:kineqg}) is integrated from $-0$ to $+0$: 
\begin{equation}
 \left[D\pd{\overline{f}}{x}\right]_2-\left[D\pd{\overline{f}}{x}\right]_1+\frac{u_2-u_1}{3}p\pd{\overline{f\rs{o}}}{p}+
 Q\rs{p}\overline{Q\rs{t}}(s)=0,
 \label{kineq2:eqgxint}
\end{equation}
the continuity condition $f_1=f_2$ (and then $\overline{f_1}=\overline{f_2}$) as well as (\ref{kineq:umova-tp}) are used. 
The expressions for the first two terms are given by differentiation of (\ref{kineq2:solxTP}) in points $+0$ and $-0$ respectively:
\begin{equation}
 \left[D\pd{\overline{f}}{x}\right]_2=\overline{f\rs{o}}u_2\beta_2= -\overline{f\rs{o}}u_2F_2/2,
\end{equation}
\begin{equation}
 \left[D\pd{\overline{f}}{x}\right]_1=\overline{f\rs{o}}u_1\beta_1= \overline{f\rs{o}}u_1F_1/2+u_1\overline{f\rs{o}},
\end{equation}
where the notations $F_1/2=\beta_1-1$, $F_2/2=-\beta_2$ are introduced. 
Then, the equation for $\overline{f\rs{o}}$ is 
\begin{equation}
 \pd{\overline{f\rs{o}}}{p}+\frac{\varsigma}{p}\overline{f\rs{o}}={q}\delta(p-p\rs{i})\overline{Q\rs{t}}
 \label{kineq2:eqgo}
\end{equation}
where
\begin{equation}
 {q}=\frac{\eta n_1u_1}{4\pi p\rs{i}^2p}\frac{3}{u_1-u_2} 
\label{kineq2:nonunitermI}
\end{equation}
and
\begin{equation}
 \varsigma=s\rs{f}+\frac{3}{2}\frac{u_1F_1+u_2F_2}{u_1-u_2}
\label{kineq2:spectrindI}
\end{equation}
is the spectral index of the function $\overline{f\rs{o}}$, i.e. $\overline{f\rs{o}}\propto p^{-\varsigma(p)}$. The term 
\begin{equation}
 s\rs{f}=\frac{3u_1}{u_1-u_2}
\label{kineq2:spectrindsf}
\end{equation}
is the spectral index of the stationary distribution function $\mathrm{f}\rs{o}(p)\equiv f\rs{o}(t\!=\!\infty,p)$. 
The general solution of the inhomogeneous equation (\ref{kineq2:eqgo}) is 
\begin{equation}
 \overline{f\rs{o}}(s,p)=\left(\overline{C}+{q}(p\rs{i})\overline{Q\rs{t}}\right)
 \exp\left[-\int_{p\rs{i}}^{p}\varsigma(s,p')\frac{dp'}{p'}\right],
\label{kineq2:generalsol}
\end{equation}
where $\overline{C}(s)$ is an arbitrary function. 

This solution may be rewritten as \citep{Drury-1983}: 
\begin{equation}
 \overline{f\rs{o}}(s,p)=\mathrm{f}\rs{o}(p)\overline{Q\rs{t}}(s)\overline{\varphi\rs{o}}(s,p),
 \label{kineq2:solgshortgen}
\end{equation}
where $\mathrm{f}\rs{o}(p)$ is the solution of the stationary equation, i.e. Eq.~(\ref{kineq:kineq}) with $\partial f/\partial t=0$:
\begin{equation}
 \mathrm{f}\rs{o}(p)=\frac{\eta n_1}{4\pi p\rs{i}^3}\frac{3u_1}{u_1-u_2}\left(\frac{p}{p\rs{i}}\right)^{-s\rs{f}},
 \label{kineq2:stationarysol}
\end{equation}
and 
\begin{equation}
 \overline{\varphi\rs{o}}(s,p)={\exp\left(-h\rs{o}(s,p)\right)},
\end{equation}
\begin{equation}
 h\rs{o}(s,p;p\rs{i})=\frac{3}{2}\int_{p\rs{i}}^{p}\frac{u_1F_1(s,p')+u_2F_2(s,p')}{u_1-u_2}\frac{dp'}{p'}.
\end{equation}
The arbitrary function $\overline{C}$ was set to zero in order to resemble the known expressions \citep[e.g.][]{Drury-1983,Blasi-2002} for the stationary solution $\mathrm{f}\rs{o}(p)$.

The distribution function $f\rs{o}(t,p)$ is given by the inverse Laplace transform (\ref{kineq2:Laplint5}) of $\overline{f\rs{o}}(s,p)$:
\begin{equation}
 f\rs{o}(t,p)=\mathrm{f}\rs{o}(p)\int\limits_{0}^{t} Q\rs{t}(t-t') \varphi\rs{o}(t') dt'
\label{kineq2:solfTPQ}
\end{equation}
where $\varphi\rs{o}(t)$ is the inverse Laplace transform of $\exp[-h\rs{o}(s)]$ and $Q\rs{t}(t)$ represents variation of the injection efficiency in time. This expression generalizes the known solution to \textit{the time-dependent injection} (some its effects on the particle spectrum are discussed in Sect.~\ref{kineq2:discussion}).

\subsection{Function $\varphi\rs{o}(t)$}
\label{kineq2:sectvarphio}

If injection is continuous $Q\rs{t}(t)=1$ then the distribution function at the shock is \citep{Drury-1983}
\begin{equation}
 f\rs{o}(t,p)=\mathrm{f}\rs{o}(p)\int_{0}^{t}\varphi\rs{o}(t')dt'.
 \label{kineq2:solfTP}
\end{equation}

If $s=0$ then $h(0)=0$. The relation
\begin{equation}
 \int_{0}^{\infty}e^{-ts}\varphi\rs{o}(t)dt=\exp[-h\rs{o}(s)],
 \label{kineq2:Mt}
\end{equation}
written for $s=0$ shows that $\varphi\rs{o}(t)$ is normalized to unity \citep{Drury-1983}:
\begin{equation}
 \int_{0}^{\infty}\varphi\rs{o}(t)dt=1.
\end{equation}
It is obvious now that the stationary solution comes from Eq.~(\ref{kineq2:solfTP}) in the limit $t\rightarrow \infty$. 

The function $\varphi\rs{o}$ allows one to derive expressions for the average acceleration time in the test-particle limit \citep{Drury-1983} as well as its generalizations: to the spatially variable diffusion coefficients \citep{Drury-1991}, to the presence of the free escape boundary upstream and downstream \citep{Ostr-Schlick-1996}, to the non-linear acceleration regime \citep{Blasi-et-al-2007}. 

We introduce notations
\begin{equation}
 h_{\mathrm{o}i}(s,p;p\rs{i})=\frac{3}{2}\int_{p\rs{i}}^{p}\frac{u_iF_i(s,p')}{u_1-u_2}\frac{dp'}{p'},
\label{kineq2:hoi}
\end{equation}
and $\varphi_{\mathrm{o}i}(t)$ as the inverse Laplace transform of $\exp(-h_{\mathrm{o}i}(s))$. 
Then the inverse Laplace transform of $\exp(-h\rs{o})=\exp(-h_{\mathrm{o}1})\cdot\exp(-h_{\mathrm{o}2})$ is 
\begin{equation}
 \varphi\rs{o}(t)=\int_{0}^{t}\varphi\rs{o1}(t')\varphi\rs{o2}(t-t')dt'.
\label{kineq2:genphi}
\end{equation} 
Eq.~(\ref{kineq2:genphi}) yields the function $\varphi\rs{o}(t)$ \textit{without limitations on the relation between $t_1$ and $t_2$}.\footnote{\citet{Forman-Drury-1983} have considered the case $t_1\gg t_2$. \citet{Toptygin-1980} has derived $\varphi\rs{o}(t)$ for $t_1=t_2$. Note that $\exp(-h\rs{o2})\rightarrow 1$ in the case $t_1\gg t_2$ and integration in (\ref{kineq2:genphi}) is not actually needed: the inverse Laplace transform of $\overline{\varphi_{\mathrm{o}}}$ is just $\varphi_{\mathrm{o}}=\varphi_{\mathrm{o}1}$. In the similar fashion, $\varphi_{\mathrm{o}}$ for $t_1=t_2$ is given by (\ref{kineq2:t1phi}) where $A_1$ should be changed to $A_1+A_2$.}

The inverse Laplace transform $\varphi_{\mathrm{o}i}$ is obtained by \citet{Forman-Drury-1983} for the diffusion coefficient of the form $D=D\rs{*}p^{\alpha}$ where $D\rs{*}$ is constant. 
We write $F_1$ and $F_2$ introduced above as
\begin{equation} 
 F_i=(1+st_i)^{1/2}-1, \quad t_i=4D_iu_i^{-2}.
 \label{kineq2:defFi}
\end{equation}
After integration in (\ref{kineq2:hoi}), and for $p\rs{o}\ll p$, one has
\begin{equation}
\begin{array}{ll}
 &\exp(-h_{\mathrm{o}i}(s))= 
 \\ \\=&\displaystyle
 \left[\frac{(1+st_i)^{1/2}+1}{2}\right]^{A_i} \exp\left[-A_i\left((1+st_i)^{1/2}-1\right)\right],
\end{array}
\label{kineq2:hsb}
\end{equation}
with
\begin{equation} 
 A_1=\frac{3\sigma}{(\sigma-1)\alpha},
 \quad
 A_2=\frac{3}{(\sigma-1)\alpha}
\end{equation}
where $\sigma=u_1/u_2$ (the value $\sigma=4$ is used for plots in the present paper). 
We consider the values of $\alpha$ providing $A_1$ to be integer\footnote{Typical diffusion coefficients satisfy this condition: $D\propto p$ -- Bohm diffusion or if the turbulence is generated by the accelerated particles with the spectrum $f(p)\propto p^{4}$; $D\propto p^{1/2}$ for diffusion in the medium with the Kraichnan turbulence spectrum; $D\propto p^{1/3}$ in the Kolmogorov turbulence \citep[e.g.][]{Amato-Blasi-2006,Blasi-2010}. If $A_1$ and $A_2$ are not integer then the solution may be written in terms of the parabolic cylinder functions \citep{Forman-Drury-1983}.}. 
Eqs.~(\ref{kineq2:binrozkl}), (\ref{kineq2:Laplint2}), (\ref{kineq2:Laplint3}) and (\ref{kineq2:Laplint4}) in the Appendix yield  \citep{Forman-Drury-1983}:
\begin{equation} 
\begin{array}{ccc}\displaystyle
 \varphi_{\mathrm{o}i}(t)&=&\displaystyle
 \frac{e^{A_i}}{2^{A_i+1}t_i\sqrt{\pi}}\frac{\exp\displaystyle\left[-\tau-{A_i^2}/{(4\tau)}\right]}{\tau}
 \\ \\ &\times& \displaystyle
 \sum_{m=0}^{A_i}C_{m}^{A_i}\left[\frac{1}{2\tau^{1/2}}\right]^m\mathrm{H}_{m+1}\left(\frac{A_i}{2\tau^{1/2}}\right),
\end{array}
\end{equation}
where $\tau=t/t_i$, $\mathrm{H}_m(x)$ -- Hermite polinomial. 

This expression may be simplified. Namely, with Eq.~(\ref{kineq2:HermCmA}), it may be reduced to
\begin{equation} 
 \varphi_{\mathrm{o}i}(t)=\frac{e^{2A_i}}{2^{2A_i+1}t_i\sqrt{\pi}}\frac{e^{-\xi(\tau)^2}}{\tau^{A_i/2+1}}
 \left(\mathrm{H}_{A_i+1}\left(\xi\right)
 -{2\tau^{1/2}}\mathrm{H}_{A_i}\left(\xi\right)
 \right),
 \label{kineq2:t1phi}
\end{equation}
where $\xi(\tau)=\tau^{1/2}+A_i/(2\tau^{1/2})$. 

\subsection{Function $f(t,x,p)$}
\label{kineq2:sectftxpappI}

The solution $f(t,x,p)$ before and behind the shock may be found by the transform (\ref{kineq2:solinvlapldef}). 
It follows from Eqs.~(\ref{kineq2:solxTP}) and (\ref{kineq2:solgshortgen}) that 
\begin{equation}
 \overline{f}(s,x,p)=\mathrm{f}\rs{o}(p)\overline{Q\rs{t}}(s)\overline{\varphi\rs{o}}(s,p)
 \overline{\varphi_{\mathrm{x}}}(s,x,p)
\label{kineq2:gensolfxLapl}
\end{equation}
where $\overline{\varphi_{\mathrm{x}}}$ 
is the exponential term in (\ref{kineq2:solxTP}). 
With the use of the property (\ref{kineq2:Laplint5}), one may derive \textit{different representations of $f(t,x,p)$} depending on how to group the terms in the expression (\ref{kineq2:gensolfxLapl}).

Namely, we may have expression which relates $f(t,x,p)$ and the distribution at the shock $f\rs{o}(t,p)$. 
Applying the transforms (\ref{kineq2:Laplint3}) and (\ref{kineq2:Laplint4}) with $n=1$ to 
\begin{equation}
 \overline{f}(s,x,p)=
 \overline{\varphi_{\mathrm{x}}}(s,x,p)
 \overline{f\rs{o}}(s,p)
\label{kineq2:invtrsol}
\end{equation}
we derive the solution in the form 
\begin{equation}
 {f}(t,x,p)=
 \int\limits_{0}^{t}\varphi_{\mathrm{x}}(t-t',x,p)
 {f\rs{o}}(t',p)dt'
\label{kineq2:gensolapII}
\end{equation}
with 
\begin{equation}
 \varphi_{\mathrm{x}}(t,x,p)=
 \frac{|x|}{(4\pi D_i t)^{1/2}t}\exp\left\{-\frac{(x-u_it)^2}{4D_it}\right\}.
\label{kineq2:varphix}
\end{equation}
(We will note the physical meaning of $\varphi_{\mathrm{x}}$ later, in Sect.~\ref{kineq2:physmeaning}.)
If $x=0$ then $\overline{\varphi_{\mathrm{x}}}=1$ and $f(t,0,p)=f\rs{o}(t,p)$. 

Another possibility is to derive expression relating $f(t,x,p)$ with the stationary solution $\mathrm{f}\rs{o}(p)$: 
\begin{equation}
 f(t,x,p)=\mathrm{f}\rs{o}(p)\int_{0}^{t}Q\rs{t}(t-t',p)\varphi(t',x,p)dt',
\label{kineq2:solfx2}
\end{equation}
where
\begin{equation}
 \varphi(t,x)={\cal L}^{-1}\{\exp(-h\rs{o}(s)+u_ix\beta_i(s)/D_i)\},
\end{equation}
and therefore 
\begin{equation}
 \varphi(t,x)=\int_{0}^{t}\varphi\rs{o}(t')\varphi_{\mathrm{x}}(t-t',x)dt'.
\label{kineq2:genphixF}
\end{equation} 
If the injection is steady-state ($Q\rs{t}=1$), in the stationary case ($t\rightarrow\infty$), Eq.~(\ref{kineq2:solfx2}) results in Eqs.~(\ref{kineq2:stacsolx-up})-(\ref{kineq2:stacsolx-dw}). Thus the function $\varphi(t,x)$ is normalized to
\begin{equation}
 \int\limits_{0}^{\infty}\varphi(t,x)dt=\left\{
  \begin{array}{ll}
	 \exp\left(u_1x/D_1\right)&x<0\\
	 1&x>0
	\end{array}
 \right. .
\label{kineq2:phinorm}
\end{equation} 

Note, that the representations (\ref{kineq2:gensolapII}) and (\ref{kineq2:solfx2}) are derived \textit{without assumptions $D\propto p^\alpha$ and $p\rs{o}\ll p$} used in Sect.~\ref{kineq2:sectvarphio}. 

The third approach to derive $f(t,x,p)$ is considered by \citet{Forman-Drury-1983}. It uses the same algorithm as for $\varphi\rs{o}$ in Sect.~\ref{kineq2:sectvarphio} and thus assumes $D\propto p^\alpha$ and $p\rs{o}\ll p$. It relates $f(t,x,p)$ with the stationary solution $\mathrm{f}(x,p)$, Eqs.~(\ref{kineq2:stacsolx-up})-(\ref{kineq2:stacsolx-dw}): 
\begin{equation}
 f(t,x,p)=\mathrm{f}(x,p)\int_{0}^{t}Q\rs{t}(t-t',p)\psi(t',x,p)dt',
\label{kineq2:solfx2b}
\end{equation}
where
\begin{equation}
 \psi(t,x)={\cal L}^{-1}\{\exp(-h(s,x))\},
\end{equation}
\begin{equation}
 h(s,x)=h\rs{o}-\frac{u_1x}{D_1}(\beta_1-1)=h\rs{o1}+h\rs{o2}+F_1B_1, \quad x<0,
\label{kineq2:h1}
\end{equation}
\begin{equation}
 h(s,x)=h\rs{o}-\frac{u_2x}{D_2}\beta_2=h\rs{o1}+h\rs{o2}+F_2B_2, \quad x>0,
\label{kineq2:h2}
\end{equation}
and $B_i=u_i|x|/(2D_i)$.
Comparing Eqs.~(\ref{kineq2:solfx2}) and (\ref{kineq2:solfx2b}), we see that $\psi(t,x)$ is normalized to unity everywhere (as for $x>0$ as for $x<0$) while $\varphi(t,x)$ in a half-space $x>0$ only, Eq.~(\ref{kineq2:phinorm}).

Inspecting the structure of Eqs.~(\ref{kineq2:h1}) and (\ref{kineq2:h2}) 
and grouping terms as $\exp[-(h\rs{o1}+F_1B_1)]\exp(-h\rs{o2})$, 
we obtain:
\begin{equation}
 \psi(t,x)=\int_{0}^{t}\varphi\rs{1}(t',x)\varphi\rs{o2}(t-t')dt', \qquad x<0,
\end{equation} 
\begin{equation}
 \psi(t,x)=\int_{0}^{t}\varphi\rs{o1}(t')\varphi\rs{2}(t-t',x)dt', \qquad x>0
\end{equation} 
where $\varphi_i$ are the inverse Laplace transforms of $\exp[-(h_{\mathrm{o}i}+F_iB_i)]$ and $\varphi_{\mathrm{o}i}$ are given by (\ref{kineq2:t1phi}).

We may find $\varphi_i(t,x)$ in the same way as $\varphi_{\mathrm{o}i}(t)$ in Sect.~\ref{kineq2:sectvarphio}. 
Namely, applying binomial decomposition (\ref{kineq2:binrozkl}) in respect to $((1+st_i)^{1/2}+1)^{A_i}$, using the shift rule (\ref{kineq2:Laplint3}) in respect to $(t_i^{-1}+s)$ and then the inverse Laplace transform (\ref{kineq2:Laplint4}) to each summand of the decomposition, we derive for $\varphi_i$ expression analogous to Eq.~(\ref{kineq2:t1phi}) with $e^{2(A_i+B_i)}$ instead of $e^{2A_i}$ and $\xi(\tau,x,p)=\tau^{1/2}+(A_i+B_i)/(2\tau^{1/2})$.
The distribution $\varphi\rs{o}(t)$ comes from $\varphi_i(t,x)$ with obvious substitution $x=0$. 

We would like to note, that \citet{Forman-Drury-1983} considers actually the case $t_1\gg t_2$; then the expressions for $\psi(t,x,p)$ for both regions $x<0$ and $x>0$ are simpler in their approach: the expressions do not contain integration (because $\exp(-h\rs{o2})\rightarrow 1$) and are given just by Eq.~(\ref{kineq2:t1phi}) with $e^{2(A_1+B_i)}$ instead of $e^{2A_1}$ and  $\xi(\tau,x,p)=\tau^{1/2}+(A_1+B_i)/(2\tau^{1/2})$. 

\section{Approach II. Conjugation problem with Laplace transform}
\label{kineq2:kineqII}

One alternative approach to solve the equation (\ref{kineq:kineq}) consists in splitting this equation into few separate equations, and in applying the Laplace transform to equations which are simpler than the original Eq.~.(\ref{kineq:kineq}). The splitting of the diffusion-convection equation into equations for $x<0$, $x>0$ and $x=0$ 
is the typical approach in solving the stationary problem \citep[e.g.][]{Drury-1983,Blasi-2002}.

From the mathematical point of view, the task to solve Eq.~(\ref{kineq:kineq}) 
may be formulated as the conjugation problem for the linear parabolic equation of the second order with discontinuous coefficients: 
\begin{eqnarray}
 \displaystyle\pd{f}{t}-\pd{}{x}\left[D_1\pd{f}{x}\right]+u_1\pd{f}{x}=0, \quad x<0,\label{kineq2:conjtask1}\\
 \displaystyle\pd{f}{t}-\pd{}{x}\left[D_2\pd{f}{x}\right]+u_2\pd{f}{x}=0, \quad x>0,\label{kineq2:conjtask2}\\
 f(0,x,p)=0, \label{kineq2:conjtask3}\\
 f_1(t,0,p)=f_2(t,0,p)\equiv f\rs{o}(t,p),  \label{kineq2:conjtask4}\\
 \displaystyle\left[D\pd{f}{x}\right]_2-\left[D\pd{f}{x}\right]_1+
 \frac{u_2-u_1}{3}p\pd{f\rs{o}}{p}+Q\rs{t}(t)Q\rs{p}(p)=0 \label{kineq2:conjtask5}
\end{eqnarray}
where, again, the index `1' refers to $x=0^-$, the index `2' to $x=0^+$ and the index `o' to $x=0$. 
The conjugation (matching) condition (\ref{kineq2:conjtask5}) is derived by integration of Eq.~(\ref{kineq:kineq}) from $x=0^-$ to $x=0^+$ under assumption that $f$, $\partial f/\partial t$, $\partial f/\partial p$ are continuous through the point $x=0$ at any time.

\subsection{Solving the parabolic conjugation problem}

The fundamental solutions of the heat conduction equations (\ref{kineq2:conjtask1}) and (\ref{kineq2:conjtask2}) are 
\begin{equation}
 g_i(t,x,\chi)=\frac{1}{(4\pi D_i t)^{1/2}}\exp\left\{-\frac{(x-u_it-\chi)^2}{4D_it}\right\},
\label{kineq2:solheat}
\end{equation}
where $\chi$ is a real variable and $D_i$ are spatially constant in their domains. 

We shall look for the solution of the conjugation problem (\ref{kineq2:conjtask1})-(\ref{kineq2:conjtask5}) in the form of the parabolic simple-layer potentials 
\begin{equation}
 f(t,x,p)=\int\limits_{0}^{t}g_i(t-\tau,x,0)V_i(\tau,p)d\tau
\label{kineq2:soleqbase}
\end{equation}
where $V_i(\tau,p)$ are unknown functions to be determined from Eqs.~(\ref{kineq2:conjtask4})-(\ref{kineq2:conjtask5}). 

Substitution (\ref{kineq2:conjtask4}) with (\ref{kineq2:soleqbase}) yields the first equation for $V_i$:
\begin{equation}
 \int\limits_{0}^{t}g_1(t-\tau,0,0)V_1(\tau,p)d\tau=\int\limits_{0}^{t}g_2(t-\tau,0,0)V_2(\tau,p)d\tau;
\label{kineq2:eqV1}
\end{equation}
note, that we used $x=0$ here. 

Dealing with the condition (\ref{kineq2:conjtask5}), we use the expression for the simple-layer potential jump which, in our case, is 
\begin{equation}
 \pd{f}{x}=(-1)^{i+1}\frac{V_i}{2D_i}+\int\limits_{0}^{t}\pd{g_i(t-\tau,0,0)}{x}V_i(\tau,p)d\tau.
\label{kineq2:jump1}
\end{equation}
We have, from (\ref{kineq2:solheat}), that 
\begin{equation}
 \pd{g_i(t-\tau,0,0)}{x}=\frac{u_i}{2D_i}g_i(t-\tau,0,0).
\label{kineq2:jump2}
\end{equation}
Now, the second equation for $V_i$ follows from Eq.~(\ref{kineq2:conjtask5}), with the use of (\ref{kineq2:jump1}), (\ref{kineq2:jump2}): 
\begin{eqnarray}
 V_1(t,p)+V_2(t,p)+u_1 \int\limits_{0}^{t}g_1(t-\tau,0,0)V_1(\tau,p)d\tau 
 \nonumber \\ \displaystyle
 -u_2\int\limits_{0}^{t}g_2(t-\tau,0,0)V_2(\tau,p)d\tau
 \nonumber \\ \displaystyle
 =\frac{2(u_2-u_1)}{3}p\pd{f\rs{o}}{p}+2Q\rs{t}Q\rs{p}
\label{kineq2:eqV2}
\end{eqnarray}

Thus, we derived the system of equations (\ref{kineq2:eqV1}) and (\ref{kineq2:eqV2}) for unknown functions $V_1$ and $V_2$ where the first equation (\ref{kineq2:eqV1}) is the Volterra integral equation of the first kind and the second one (\ref{kineq2:eqV2}) is the Volterra integral equation of the second kind. There is unknown function $\partial f\rs{o}(t,p)/\partial p$ in the right-hand side of Eq.~(\ref{kineq2:eqV2}). Let us find $f\rs{o}$ before solving the system (\ref{kineq2:eqV1}), (\ref{kineq2:eqV2}).

\subsection{Function $f\rs{o}(t,p)$}

Both the integrals in Eq.~(\ref{kineq2:eqV1}) are equal to 
\begin{equation}
 \int\limits_{0}^{t} g_i(t-\tau,0,0)V_i(\tau,p)d\tau=f\rs{o}(t,p),
\label{kineq2:eqfo1}
\end{equation}
due to the continuity of the distribution function, (\ref{kineq2:conjtask4}). 
With this relation, Eq.~(\ref{kineq2:eqV2}) becomes
\begin{eqnarray}
 V_1(t,p)+V_2(t,p)=(u_2-u_1)\left(\frac{2}{3}p\pd{f\rs{o}}{p}+f\rs{o}\right)+2Q\rs{t}Q\rs{p}.
\label{kineq2:eqfo2}
\end{eqnarray}
In order to obtain the equation for $f\rs{o}$, we apply the Laplace transform to (\ref{kineq2:eqfo1}) and (\ref{kineq2:eqfo2}). 

The first one, Eq.~(\ref{kineq2:eqfo1}), with the use of the convolution property (\ref{kineq2:Laplint5}) transforms to 
\begin{equation}
 \overline{g_i}(s,0,0)\ \overline{V_i}(s,p)=\overline{f\rs{o}}(s,p)
\label{kineq2:eqfo3}
\end{equation}
Eq.~(\ref{kineq2:solheat}) yields 
\begin{equation}
 \overline{g_i}(s,0,0)=\left(4D_i\right)^{-1/2}\left(s+u_i^2/4D_i\right)^{-1/2}.
\label{kineq2:eqfo3b}
\end{equation}
These two relations allow us to find 
\begin{equation}
 \overline{V_i}(s,p)=\left(4D_i\right)^{1/2}\left(s+u_i^2/4D_i\right)^{1/2}\overline{f\rs{o}}(s,p)
\label{kineq2:eqfo3c}
\end{equation}
and their sum $\overline{V_1}+\overline{V_2}$. 

The second one, Eq.~(\ref{kineq2:eqfo2}), after the Laplace transform, gives another equation for the sum $\overline{V_1}+\overline{V_2}$: 
\begin{equation}
 \overline{V_1}+\overline{V_2}=
 (u_2-u_1)\left(\frac{2}{3}p\pd{\overline{f\rs{o}}}{p}+\overline{f\rs{o}}\right)+2\overline{Q\rs{t}}(s)Q\rs{p}.
\end{equation}

Equating the two expressions for $\overline{V_1}+\overline{V_2}$, we derive the differential equation for $\overline{f\rs{o}}(s,p)$ which is exactly the same as Eqs.~(\ref{kineq2:eqgo}). Its solution gives the function $f\rs{o}$, as it is shown in Sect.~\ref{kineq2:sectfotp} and \ref{kineq2:sectvarphio}: 
\begin{equation}
 f\rs{o}(t,p)=\mathrm{f}\rs{o}(p)
 \int\limits_{0}^{t} dt' Q\rs{t}(t-t') 
 \int_{0}^{t'} dt'' \varphi\rs{o1}(t'',p)\varphi\rs{o2}(t'-t'',p).
\label{kineq2:gensol}
\end{equation} 

\subsection{Function $f(t,x,p)$}
\label{kineq2:physmeaning}

The function $f(t,x,p)$ may be obtained by substitution (\ref{kineq2:soleqbase}) with expression for $V_i$. 
In order to have the expression for $V_i$, we use (\ref{kineq2:eqfo3b}) in (\ref{kineq2:eqfo3c}) and write 
\begin{equation}
 \overline{V_i}(s,p)=4D_i\left(s+u_i^2/4D_i\right)\overline{g_i}(s,0,0)\overline{f\rs{o}}(s,p).
\label{kineq2:eqfo3d}
\end{equation}
Inverting this, we come to
\begin{equation}
 {V_i}(t,p)=4D_i\left(\pd{}{t}+\frac{u_i^2}{4D_i}\right)\int\limits_{0}^{t}{g_i}(t',0,0){f\rs{o}}(t-t',p)dt'.
\label{kineq2:Via}
\end{equation}

An alternative possibility to derive $f(t,x,p)$, without the need to know $V_i$, is to consider the Laplace transform of Eq.~(\ref{kineq2:soleqbase}):
\begin{equation}
 \overline{f}(s,x,p)=\overline{g_i}(s,x,0)\overline{V_i}(s,p)
 \label{kineq2:laplf(xp)}
\end{equation}
and to represent ${g_i}(s,x,0)$, given by (\ref{kineq2:solheat}), as 
\begin{equation}
\begin{array}{ccl}
 g_i(t,x,0)
 &=&\displaystyle
 \exp\left(\frac{u_ix}{2D_i}\right)
 \exp\left(-\frac{u_i^2}{4D_i}\,t\right)
 \\ \\ &\times&\displaystyle
 \left(\frac{1}{4\pi D_it}\right)^{1/2}
 \exp\left(-\frac{x^2}{4D_it}\right).
\end{array}
\end{equation}
Then we have i) to apply the Laplace transform to this $g_i$ (the properties to be used are (\ref{kineq2:Laplint3}) and (\ref{kineq2:Laplint4}) with $n=0$), ii) to express $\overline{V_i}$ from (\ref{kineq2:eqfo3}) with (\ref{kineq2:eqfo3b}) and iii) to substitute these $\overline{g_i}$ and $\overline{V_i}$ into Eq.~(\ref{kineq2:laplf(xp)}). After these steps we have that 
\begin{equation}
 \overline{f}(s,x,p)=
 \overline{\varphi_{\mathrm{x}}}(s,x,p)
 \overline{f\rs{o}}(s,p)
\label{kineq2:invtrsoli}
\end{equation}
where 
\begin{equation}
 \overline{\varphi_{\mathrm{x}}}(s,x,p)=\exp\left(\frac{u_ix}{2D_i}\right)
 \exp\left[-\frac{u_i|x|}{2D_i}\left(1+\frac{4sD_i}{u_i^2}\right)^{1/2}\right]
\end{equation}
which is the same as $\overline{\varphi_{\mathrm{x}}}$ used in Sect.~\ref{kineq2:sectftxpappI}.
Now, applying the inverse Laplace transform to (\ref{kineq2:invtrsoli}), we come to the solution $f(t,x,p)$ which is the same as (\ref{kineq2:gensolapII}). 

Note that $\varphi_{\mathrm{x}}$ is in fact (cf. Eq.~\ref{kineq2:varphix})
\begin{equation}
 \varphi_{\mathrm{x}}(t,x)=\frac{|x|}{t}g_i(t,x,0).
\end{equation}
This demonstrates the close relation of the time-dependent acceleration problem to the fundamental solution of the heat conduction equation and reveals the physical meaning of $\varphi_{\mathrm{x}}$:  
shift of the distribution in space with velocity $x/t$ and its spread in accordance to $g_i$.

\section{Approach III. Conjugation problem without Laplace transform}
\label{kineq2:kineqIII}

The third approach to the time-dependent acceleration problem deals again with the conjugation equations (\ref{kineq2:conjtask1})-(\ref{kineq2:conjtask5}) but without the Laplace transform. Generally speaking, in this way we may overcome the conditions $\alpha=\mathrm{const}$ and $p\gg p\rs{i}$ used during the inverse Laplace transform in Sect.~\ref{kineq2:sectvarphio}. The former possibility is important in cases where the dependence of the diffusion coefficient $D$ on the particle momentum differs from the power law. The later possibility could be relevant for small particle momenta which are not well above the injection momentum $p\rs{i}$ 
(this is almost unimportant in the astrophysical environments). 

Like in the previous section, the solutions of the heat equations (\ref{kineq2:conjtask1}) and (\ref{kineq2:conjtask2}) are given by (\ref{kineq2:solheat}). We consider the Volterra integral equation of the first kind (\ref{kineq2:eqV1}). In order to regularize it, we consider the operator $\hat {\cal E}_i$ which maps according to the rule
\begin{equation}
 \hat {\cal E}_i(t) \psi(t)=4D_i\left(\pd{}{t}+\frac{u_i^2}{4D_i}\right)\int\limits_{0}^{t}{g_i}(t-t',0,0)\psi(t')dt'.
\end{equation}
Applying it to both sides of (\ref{kineq2:eqfo1}), we obtain expressions (\ref{kineq2:Via}) for $V_1$ and $V_2$ which, if the condition $f\rs{o}(0,p)=0$ holds, may be written as
\begin{equation}
 {V_i}(t,p)=4D_i\int\limits_{0}^{t}{g_i}(t',0,0)\left(\pd{}{t}+\frac{u_i^2}{4D_i}\right){f\rs{o}}(t-t',p)dt'.
\label{kineq2:Viab}
\end{equation}

The function $f_i(t,x,p)$ may be found by substitution Eq.~(\ref{kineq2:soleqbase}) with (\ref{kineq2:Via}) or (\ref{kineq2:Viab}). 

An \textit{equation for the non-stationary distribution function at the shock} $f\rs{o}$ comes from Eq.~(\ref{kineq2:eqfo2}):
\begin{equation}
 \pd{f\rs{o}}{p}+\frac{3}{2}\frac{f\rs{o}}{p}+\frac{3}{2}\frac{V_1+V_2}{u_1-u_2}\frac{1}{p}
 ={q}\delta(p-p\rs{i}){Q\rs{t}}.
\label{kineq2:eqfonoLt2}
\end{equation}
It is worth to note that this equation is derived without Laplace transform. 

The correctness of the equation may be demonstrated by converting it to the equation (\ref{kineq2:eqgo}) for the Laplace transform $\overline{f\rs{o}}(s,p)$. The way to do this is following: i) compare (\ref{kineq2:eqfo3c}) and (\ref{kineq2:defFi}) and note that $\overline{V_i}=\overline{f\rs{o}}u_i(F_i+1)$, ii) apply the Laplace transform to (\ref{kineq2:eqfonoLt2}) and substitute it with this relation between $\overline{V_i}$ and $F_i$. 

Introducing the variables $W_i$ defined by relations $V_i=f\rs{o}u_i(W_i+1)$ with $V_i$ given by (\ref{kineq2:Viab}), 
we come to the integro-differential equation for $f\rs{o}$:
\begin{equation}
 \pd{f\rs{o}}{p}+\frac{s\rs{t}}{p}f\rs{o}={q}\delta(p-p\rs{i}){Q\rs{t}}
\label{kineq2:eqfonoLt}
\end{equation}
where 
\begin{equation}
 s\rs{t}=s\rs{f}+\frac{3}{2}\frac{u_1W_1+u_1W_2}{u_1-u_2},
\end{equation}
$q$ is given by (\ref{kineq2:nonunitermI}) and $s\rs{f}$ by (\ref{kineq2:spectrindsf}).
Assuming $W_i$ are known, the solution is: 
\begin{equation}
 f\rs{o}=\mathrm{f}\rs{o}(p)Q\rs{t}(t){\cal F}(t,p)
 \label{kineq2:solfoapprIII}
\end{equation}
with
\begin{equation}
 {\cal F}(t,p)=\exp\left[-\frac{3}{2}\int_{p\rs{i}}^{p} \frac{u_1W_1+u_2W_2}{u_1-u_2} \frac{dp'}{p'}\right] .
 \label{kineq2:eqcalF}
\end{equation}
Since ${\cal F}$ is expressed through the function $f\rs{o}$ which we are looking for, the final solution for $f\rs{o}$ may be obtained by the method of successive approximations.

The limit of ${\cal F}(t)$ is zero for $t\rightarrow 0$ due to (\ref{kineq2:umova1}) and is unity for $t\rightarrow \infty$ since $f(\infty,x,p)=\mathrm{f}(x,p)$ by definition.

Comparing (\ref{kineq2:solfoapprIII}) and (\ref{kineq2:solfTPQ}), we note that 
\begin{equation}
 Q\rs{t}(t){\cal F}(t)=\int_{0}^{t} Q\rs{t}(t-t') \varphi\rs{o}(t') dt'.
\end{equation}
The relation between ${\cal F}$ and $\varphi\rs{o}$ is simpler for continuous injection (i.e. $Q\rs{t}=1$):
${\cal F}=\int_{0}^{t}\varphi\rs{o}dt$.

It is useful for applications to note that ${\cal F}$ depends in fact on the one variable $\tau$ only which is a combination $\tau=t/t_1=tu_1^2/4D_1(p)$. Really,
\begin{equation}
 {\cal F}(\tau)=\exp\left[-\frac{3}{2}\int_{\tau}^{\tau\rs{i}} \frac{\sigma W(\tau')+W(\chi\tau')}{\sigma-1} 
  \frac{1}{\alpha}\frac{d\tau'}{\tau'}\right] 
 \label{kineq2:eqcalFtau}
\end{equation}
where $\tau\rs{i}=tu_1^2/4D_1(p\rs{i})$, $\sigma=u_1/u_2$, $\chi=t_1/t_2$, $\alpha=d\ln D(p)/d\ln p$ which may be calculated for any (differentiable) dependence of the diffusion coefficient on the particle momentum and
\begin{equation}
 W(\tau)=\frac{1}{\phi(\tau)}\int\limits_{0}^{\tau}\frac{\exp(-\tau)}{\pi^{1/2}\tau^{1/2}}
 \left(\pd{}{\tau}+1\right)\phi(\tau-\tau')d\tau'
 -1
\label{kineq2:Wdef}
\end{equation}
where we denote $\phi=Q\rs{t}{\cal F}$.
Therefore, the method of successive approximations reads
\begin{equation}
\begin{array}{l}
 \displaystyle
 {\cal F}^{\left\langle k+1\right\rangle}(\tau)=
 \\ \displaystyle
 \exp\left[-\frac{3}{2}\int\limits_{\tau}^{\tau\rs{i}} 
 \frac{\sigma W(\tau';{\cal F}^{\left\langle k\right\rangle}(\tau'))+
 W(\chi\tau';{\cal F}^{\left\langle k\right\rangle}(\chi\tau'))}{\sigma-1} 
 \frac{1}{\alpha}\frac{d\tau'}{\tau'}\right] 
\end{array}
\end{equation}
with the initial guess value 
\begin{equation}
 {\cal F}^{\left\langle 0\right\rangle}(\tau)=\int_{0}^{\tau} \varphi_{\mathrm{o}1}(\tau')d\tau' .
\end{equation}

It should be noted that this approach to solve the time-dependent equation is highly demanding from the computational point of view because each iteration increases the number of enclosed integrals.

\section{Discussion}
\label{kineq2:discussion}

\begin{figure*}
 \centering
 \includegraphics[width=18truecm]{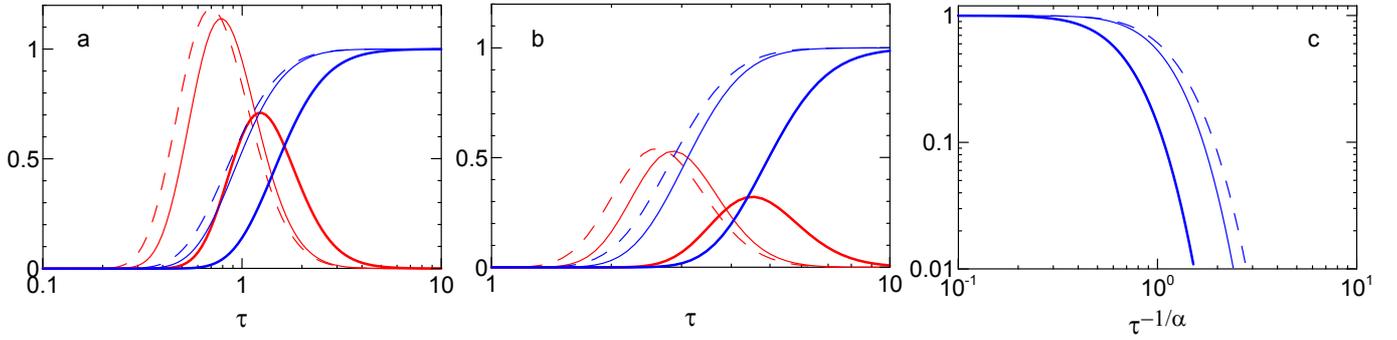}
 \caption{The function $t_1\varphi\rs{o}(\tau)$ (red lines) and integral $\int_0^\tau \varphi\rs{o}(\tau)d\tau$ (blue lines), calculated from Eq.~(\ref{kineq2:t1phi}) with assumption $t_1\gg t_2$ (then $\varphi\rs{o}=\varphi\rs{o1}$; long dashed lines) and from Eq.~(\ref{kineq2:genphi}) for $t_1/t_2=3$ (thin solid lines) and $1/3$ (thick solid lines). Plot ({\bf a}) for $\alpha=1$, plot ({\bf b}) for $\alpha=1/3$. 
Plot ({\bf c}) represents dependence of the integral on $\tau^{-1/\alpha}\propto p/p\rs{max}$ for the same cases as plot ({\bf a}). 
               }
 \label{kineq2:fig_a}
\end{figure*}
\begin{figure*}
 \centering
 \includegraphics[width=18truecm]{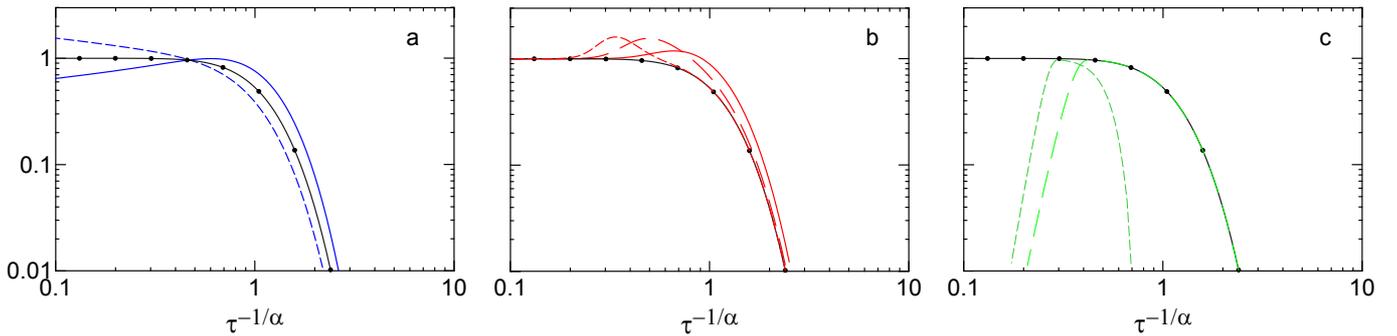}
 \caption{The integral in Eq~(\ref{kineq2:solfTPQ}) for constant injection $Q\rs{t}=1$ is shown by the black line with circles on all three plots. 
Plot {\bf a}: The same integral for decreasing or increasing injection of the form $Q\rs{t}=\tau^{b}$ with $b=-1/5$ (solid blue line), $b=1/5$ (dashed blue line).
Plot {\bf b}: The red lines corresponds to the time dependence given by Eq.~(\ref{kineq2:QtGaussian}) with $a\rs{*}=1$, $\sigma\rs{*}=0.5$ and $\tau\rs{*}=0$ (red solid line), $\tau\rs{*}=1$ (red long-dashed line) and $\tau\rs{*}=2$ (red short-dashed line). 
Plot {\bf c}: The green lines corresponds to Eq.~(\ref{kineq2:QtHeaviside}) with $\tau_1=0$ and $\tau_2=2$ (green long-dashed line) and $\tau_1=1$ and $\tau_2=3$ (green short-dashed line). 
In all cases $\alpha=1$, $t_1/t_2=3$.
               }
 \label{kineq2:fig_b}
\end{figure*}

\subsection{The solution for $t_1\gg t_2$ and the more general expression}
\label{kineq2:pmax00}

The maximum momentum of accelerated particles $p\rs{max}$ may be obtained from the expression for the average acceleration time \citep{Drury-1983}
\begin{equation}
 \left\langle t\right\rangle=\frac{3}{u_1-u_2}\int_{p\rs{i}}^{p} \left(\frac{D_1(p')}{u_1}+\frac{D_2(p')}{u_2}\right)\frac{dp'}{p'}.
 \label{kineq2:taccave}
\end{equation}
We substitute it with $D=D\rs{*}p^\alpha$, integrate and determine the maximum momentum $p\rs{max}\gg p\rs{i}$ from the equation $\left\langle t\right\rangle(p\rs{max})=t$:
\begin{equation} 
 p\rs{max}=\left(\frac{\alpha t u_1}{s\rs{f}\left(D\rs{1*}/u_1+D\rs{2*}/u_2\right)}\right)^{1/\alpha}.
 \label{kineq2:highplimit4a}
\end{equation}
This result is valid for any relation between $t_1$ and $t_2$. The expressions for $p\rs{max}$ in limits $t_1\gg t_2$ (index 1) and $t_2\gg t_1$ (index 2) follows from (\ref{kineq2:highplimit4a}): 
\begin{equation} 
 p_{\mathrm{max}i}=\left(\frac{\alpha t u_1}{s\rs{f}\left(D_{i*}/u_i\right)}\right)^{1/\alpha}.
\end{equation}
The maximum momenta are determined by the ratio of the two length-scales: of the shock motion $u_1t$ and of the particle diffusion $D_i/u_i$. 

Rewriting (\ref{kineq2:highplimit4a}) in terms of $t_{i\mathrm{*}}=4D_{i\mathrm{*}}/u_i^2$, namely,
\begin{equation} 
 p\rs{max}=\left(\frac{4\alpha t }{s\rs{f}\left(t\rs{1*}+t\rs{2*}/\sigma\right)}\right)^{1/\alpha}
\end{equation}
we see that $p\rs{max}$ shifts toward the smaller momenta with increase of $t_2$ comparing to the solution which assumes $t_1\gg t_2$. 

The function $t_1\varphi\rs{o}(\tau)$ and the integral ${\cal F}=\int_0^\tau \varphi\rs{o}(\tau)d\tau$ are shown on Fig.~\ref{kineq2:fig_a}a,b for two indexes of the diffusion coefficient. It is clear that the probability distribution $\varphi\rs{o}(\tau)$ has a peak \citep{Drury-1983}. The integral ${\cal F}$ is zero at $\tau=0$ and reaches unity with increasing $\tau$. Since $\tau\propto D_1(p)^{-1}$, the function ${\cal F}(\tau)$ represents also dependence on the particle momentum $p$. Fig.~\ref{kineq2:fig_a}c shows the integral ${\cal F}$ versus 
\begin{equation} 
 \tau^{-1/\alpha}=\frac{tu_1^2}{4D_1(p\rs{max})}\frac{p}{p\rs{max}}
\end{equation}
and therefore demonstrates the shape of the particle spectrum around $p\rs{max}$ for the steady-state injection. The \textit{shift of $p\rs{max}$ toward smaller momenta with increase of the downstream acceleration time-scale} is visible on this plot as well.

Fig.~\ref{kineq2:fig_a} compares the Drury's solution (\ref{kineq2:t1phi}) derived under assumption $t_1/t_2\gg 1$ (dashed lines) with the more general expression (\ref{kineq2:genphi}) for $t_1/t_2=3$ (thin solid lines) and $1/3$ (thick solid lines): the more significant $t_2$ the larger the difference between the lines. However, this effect is almost unimportant if the ratio $t_1/t_2$ is larger than few. The simpler Drury's formula $\varphi\rs{o}=\varphi\rs{o1}$ may then be used in order to approximate the solution. 

\subsection{The time dependent injection and the particle spectrum around $p\rs{max}$}
\label{kineq2:injteffect}

Being accelerated, particles become more energetic with time. A particle population `moves' with time along the spectrum from the lowest to the largest energies. Most of the particles, which started acceleration early, have larger energies at a given time, comparing to particles injected recently. 

Modern models of the high-energy emission from SNRs, even quite sophisticated, assume the constant injection efficiency (i.e. a fraction of particles to be accelerated). Is there any sign that shocks could be able to keep this fraction constant under different conditions and during ages? In particular, theoretical consideration of the transport of magnetic turbulence together with the particle acceleration requires the injection to be variable in time \citep{Brose-et-al-2016}. Does and how the time-dependent injection affect the particle spectrum (and therefore  their emission)?

A harder particle spectrum at the highest energies is typically considered as a sign that particle acceleration is in the effective non-linear regime. However, the variable injection in the time-depending test-particle acceleration  may also be responsible for hardening of the spectrum, if the efficiency of injection monotonically decreases (blue solid line on Fig.~\ref{kineq2:fig_b}a). In such a situation, more particles reach the high-energy end (being accelerated from early times when the injection was more effective) with respect to the less-energetic part of the spectrum (where particles injected recently with lower efficiency reside). In contrast, if the efficiency of injection is constantly increasing with time then the spectrum is relatively softer around $p\rs{max}$ (blue dashed line on Fig.~\ref{kineq2:fig_b}a). 

A more prominent bump in the particle spectrum around the largest energies is expected if a source was more effective in injection around a limited period of time at the beginning (e.g. during some time after the supernova explosion). 
We use a toy model of the continuous injection plus Gaussian 
\begin{equation}
 Q\rs{t}(\tau)=1+\frac{a\rs{*}}{\sqrt{2\pi\sigma\rs{*}^2}}\exp\left[-\frac{(\tau-\tau\rs{*})^2}{2\sigma\rs{*}^2}\right] 
 \label{kineq2:QtGaussian}
\end{equation}
in order to simulate a source which effectively injected particles around some time $\tau\rs{*}$ and is supplying them at a constant rate at other times. Red lines on Fig.~\ref{kineq2:fig_b}b demonstrate that the earlier the particles were injected the larger their momenta are at the present time, as expected. 
Efficient particle injection around $\tau\rs{*}=0$ results in a bump around $p\rs{max}$ (Fig.~\ref{kineq2:fig_b}b, solid red line). 

\begin{figure}
 \centering
 \includegraphics[width=8truecm]{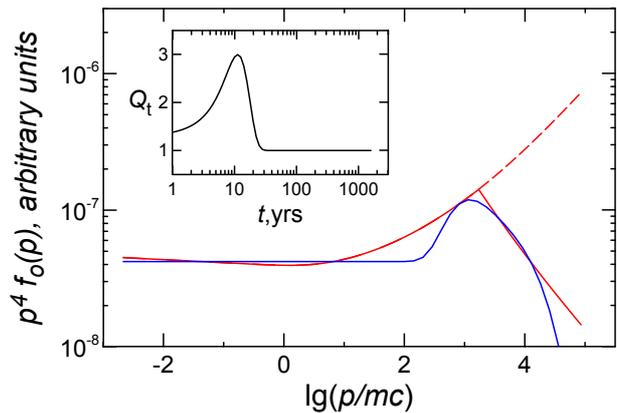}
 \caption{The spectrum of accelerated protons in RX J17.13.7--3946: the \citet{Blasi-2002} non-linear steady state solution (dashed red line), the same with the break \citep{Malkov-et-al-2005} (solid red line) and the time-dependent test-particle solution Eq.~(\ref{kineq2:solfTPQ}). {\em Inset} shows the time variation of the injection efficiency used to produce the blue line on the main plot (see text for details).
               }
 \label{kineq2:figRX}
\end{figure}
 
As an example, we consider the SNR RX J17.13.7--3946. The time-dependent injection in the test-particle limit may lead to a shape of the particle spectrum similar to the one which could be due to two effects: efficient acceleration in the non-linear regime and a spectral break due to deterioration of the particle confinement if the shock expands near regions of the weakly ionized medium \citep{Malkov-et-al-2005,Malkov-et-al-2010}. Fig.~\ref{kineq2:figRX} (red dashed line) shows the proton spectrum as it is given by the non-linear steady-state solution of \citet{Blasi-2002}. This line is plotted  for the same parameters as adopted by \citet{Malkov-et-al-2005} for SNR RX J17.13.7--3946: $p\rs{max}=10^5 mc$, the Mach number $M=80$, 
the overall shock compression $\sigma\rs{tot}=15$, injection efficiency $\eta=1.7\E{-5}$. The solid line represents the same spectrum with a break at $p=1.8\E{3}mc$ (red solid line); the spectral index of the spectrum is $\varsigma\rs{br}(p)=\varsigma(p)-1$ after the break \citep{Malkov-et-al-2005} where $\varsigma(p)$ is the index of the distribution function shown by the dashed line. The blue line is a spectrum calculated with the time-dependent test-particle solution (\ref{kineq2:solfTPQ}) with: $t\approx 1600\un{yrs}$, corresponding to the age of RX J17.13.7--3946 if it is the remnant of the supernova AD393 \citep{Wang-et-al-1997}, the shock velocity $V=1000\un{km/s}$ and the injection momentum $p\rs{i}=2.1\E{-3}mc$ \citep{Malkov-et-al-2005}, and the diffusion coefficient $D(p\rs{max})=8\E{25}\un{cm^2/s}$ with the index $\alpha=1/3$. 
The variation of the injection with time $Q\rs{t}$ is described by Eq.~(\ref{kineq2:QtGaussian}) with $a\rs{*}=10$, $\tau\rs{*}=4$ and $\sigma\rs{*}=2$ and, for parameters considered, is represented on the inset plot on Fig.~\ref{kineq2:figRX}: the injection is highest during the first 30 years after the explosion, then the particles enters the acceleration process at a steady-state, lower, rate. The time-dependent spectrum demonstrates that the hardness around the maximum momentum is similar to the nonlinear steady-state solution with the spectral break and could thus also explain the observed emission spectrum of this SNR. Note that difference between injection efficiency at the maximum and at the constant rate is rather small in this example, it is about 3 times only. 

It is worth stressing, that \textit{the particles injected during the first decades after the SN event are actually responsible for the shape of the high-energy end of the particle spectrum} (and thus of the high-energy emission) of SNR at the present time. Thus, the consideration of the variable injection could be a crucial element in models explaining the observed X-ray and gamma-ray spectra of young SNRs.

The last point in this subsection: what is the distribution function $f\rs{o}$ if the particles were injected during a limited period only and then the injection switched off? It is modeled here with a simple expression 
\begin{equation}
 Q\rs{t}(\tau)=\mathrm{H}(\tau-\tau_1)\mathrm{H}(\tau_2-\tau),
\label{kineq2:QtHeaviside}
\end{equation}
where $\mathrm{H}$ is the Heaviside step function, and the result is shown on Fig.~\ref{kineq2:fig_b}c. It is clear again that the highest energy particles are those injected at the earliest times. The low-energy cutoff is evident in the particle spectrum; it appears due to the suppression of the injection after $\tau_2$. 

\begin{figure}
 \centering
 \includegraphics[width=8.4truecm]{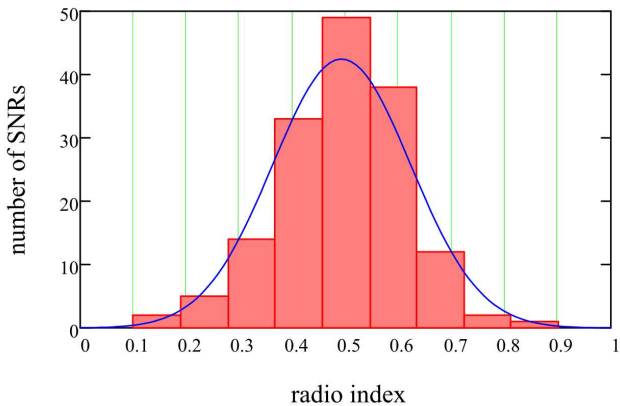}
 \caption{The distribution of the known radio spectral indices of 156 Galactic SNRs. Data are from \citet{Green-2014} catalogue. The maximum, minimum and average values of $\alpha\rs{r}$ are $0.1$, $0.9$, $0.49$ respectively, standard deviation $0.13$. 
               }
 \label{kineq2:fig_alpha}
\end{figure}

\subsection{Slope of the spectrum at the intermediate momenta}
\label{kineq2:pinter}

If injection varies only during the first decades after the supernova explosion, then it affects the spectral shape just near the high-energy cut-off. The slope of the particle spectrum at momenta much less than the cut-off momenta  remains unchanged in such a model (Fig.~\ref{kineq2:figRX}) because the injection process supplied particles at a constant rate at later times (from which particles were accelerated to smaller energies; Fig.~\ref{kineq2:figRX}). Such kind of the time dependence of the injection could be important for interpretation of the $\gamma$-ray and X-ray observations of SNRs (emission from the highest-energy particles) but not for the radio radiation. However, what if the injection varies also later on?

Electrons emitting at a radio frequency, say at $1\un{GHz}$ in the magnetic field $B\sim 30\un{\mu G}$, have energy $\sim 3\un{GeV}$. The acceleration time-scale for such electrons is of the order of a week, assuming the Bohm diffusion and the shock speed $1000\un{km/s}$. This is much less than the acceleration time of the highest-energy particles which is comparable to the age of an SNR. Therefore, the radio observations may reveal the {\it present-time} behavior of the injection efficiency. 

The test-particle shock acceleration in a {\em stationary} regime predicts that the spectral index $s\rs{f}$ of the distribution $f\rs{o}(p)$ depends on the shock compression $\sigma$ only; it is $s\rs{f}=3\sigma/(\sigma-1)$. The value $\sigma=4$ typical for the strong astrophysical shocks in a media with $\gamma=5/3$ should result in $s\rs{f}=4$ and in the spectral index of the synchrotron emission $\alpha\rs{r}=(s\rs{f}-3)/2=0.5$. 
The non-stationary consideration with the time-dependent injection could lead to deviation of the spectral index $\varsigma$, Eq.~(\ref{kineq2:spectrindI}), from the canonical value $s\rs{f}=4$ (Fig.~\ref{kineq2:fig_b}a, blue lines).

Actually, the radio observations could help us to understand how strong could be the time dependence of the injection efficiency in SNRs. Let's consider an extreme assumption, namely, that the observed spread in the radio spectral index $\alpha\rs{r}$ (Fig.~\ref{kineq2:fig_alpha}) is completely due to the variable injection. 
We take for this estimate the temporal term in (\ref{kineq2:injtermgen}) to be of the form 
\begin{equation}
 Q\rs{t}\propto t^{\beta}
\end{equation}
with constant $\beta$. In order to estimate the effect of the index $\beta$, one may calculate numerically the radio index $\alpha\rs{r}$ from the solution of the time-dependent equation:
\begin{equation}
 \alpha\rs{r}=-\frac{1}{2}\pd{\ln f\rs{o}(p)}{\ln p}-\frac{3}{2}
 \label{kineq2:defalphar}
\end{equation}
with Eq.~(\ref{kineq2:solfTPQ}) for $f\rs{o}(p)$ and Eq.~(\ref{kineq2:genphi}) for $\varphi\rs{o}$, taking the value of the index at $p\ll p\rs{max}$. We find by numerical calculations that, in order to reproduce the observed range $\alpha\rs{r}=0.1\div 0.9$, the index $\beta$ should be between $\beta=-0.7\div 0.7$ if $\alpha=1$ and between $\beta=-2.0\div 2.0$ if $\alpha=1/3$. The dependence $\alpha\rs{r}(\beta)$ is linear for other parameters fixed. 

It is remarkable that the dependence of the radio spectral index on the parameters which determine the non-stationary distribution function may be found analytically. Really, $\varphi\rs{o}$ has a peak. If we substitute Eq.~(\ref{kineq2:solfTPQ}) with the delta-function instead of $\varphi\rs{o}$ and take the derivative (\ref{kineq2:defalphar}), we obtain that
\begin{equation}
 \alpha\rs{r}=\frac{(s\rs{f}+\alpha \beta)-3}{2}.
 \label{kineq2:alpharapprox}
\end{equation}
Though this expression is approximate it appears to be close to the dependence derived numerically. 
This formula leads to an important conclusion. \textit{If the acceleration does not reach the steady state yet} (like in the young SNRs), {\it the spectral index of the accelerated particles} at the intermediate momenta $p\rs{i}\ll p\ll p\rs{max}$ (and thus the radio spectral index) \textit{is not the function of the shock compression factor $\sigma$ only} (like it is in a stationary system for the index $s\rs{f}$) but depends also on the index $\alpha$ which represent the dependence of the diffusion coefficient on the momentum, $D\propto p^{\alpha}$, and the index $\beta$ in the temporal variation of the injection efficiency, $Q\rs{t}\propto t^{\beta}$. In contrast, the time-dependent injection in the stationary regime of the particle acceleration affects the only normalization of the spectrum (through the coefficient $\eta$ in Eq.~\ref{kineq2:stationarysol}), but not the spectral index.

\section{Conclusions}
\label{kineq2:conclusions}

In the present paper, we generalized the solution of \citet{Drury-1983,Forman-Drury-1983} which describes the time-dependent diffusive shock acceleration of test-particles. The three representations of the spatial variation of the particle distribution function $f(t,x,p)$ are presented. Namely, 
Eq.~(\ref{kineq2:gensolapII}) gives $f(t,x,p)$ through $f(t,x=0,p)$, 
Eq.~(\ref{kineq2:solfx2}) yields $f(t,x,p)$ versus $f(t=\infty,x=0,p)$, 
Eq.~(\ref{kineq2:solfx2b}) relates $f(t,x,p)$ with $f(t=\infty,x,p)$.
Our generalized solution (\ref{kineq2:gensol}) for the distribution function at the shock $f\rs{o}(t,p)\equiv f(t,x=0,p)$ is valid for any ratio between the acceleration time-scales upstream and downstream of the shock and allows one to consider the time variation of the injection efficiency. 

It is shown that, if the ratio $t_1/t_2$ decreases (i.e. the significance of the downstream acceleration time grows) then the particle maximum momentum is smaller comparing to $p\rs{max}$ calculated under assumption $t_1 \gg t_2$.
The reason is visible from Eq.~(\ref{kineq2:highplimit4a}). Namely, $p\rs{max}$ is determined by the ratio between the length-scale of the shock motion and lenght-scale of the diffusion: the larger the diffusion lenght-scale the smaller the maximum momentum. 

However, if the ratio $t_1/t_2$ is larger than few then the simpler expression (\ref{kineq2:solfTPQ}) may be used for the particle distribution function $f\rs{o}(t,p)$, with $\varphi\rs{o}$ given by (\ref{kineq2:t1phi}). If, in addition, the injection is continuous and constant ($Q\rs{t}=1$), then our generalized solution becomes the same as in \citet{Drury-1983,Forman-Drury-1983}. 

The time dependence of the injection efficiency is an important factor in formation of the shape of the particle spectrum at all momenta. 

The high-energy end of the accelerated particle spectrum is formed by particles injected at the very beginning. Therefore, the temporal evolution of injection, especially during the first decades after the supernova explosion,  does affect the non-thermal spectra of young SNRs and has to be considered in interpretation of the X-ray and gamma-ray data. 

The stationary solution of the shock particle acceleration predicts that the power-law index of the cosmic ray distribution $s\rs{f}$ is determined by the shock compression only. In contrast, in young SNRs where acceleration is not presumably steady-state, this index (let's call it $s\rs{t}$ to distinguish from the stationary index $s\rs{f}$) depends also on the indexes $\alpha$ and $\beta$ in the approximate expressions for the diffusion coefficient $D\propto p^{\alpha}$ and for the temporal evolution of the injection efficiency $Q\rs{t}\propto t^{\beta}$. Namely, it is $s\rs{t}\approx s\rs{f}+\alpha\beta$. This property of the time-dependent solution could be responsible for deviation of the observed radio index from the classical value $0.5$ in some young SNRs. 

Since the acceleration times for electrons emitting at radio frequencies are very small, the observed slopes of the radio spectra could reflect the current evolution of the injection in SNRs. 

\section*{Acknowledgements}

This work is partially funded by the PRIN INAF 2014 grant `Filling the gap between supernova explosions and their remnants through magnetohydrodynamic modeling and high performance computing'.



\appendix
\section[]{Necessary identities}
\label{kineq2:app1}

In the main text of the paper, the binomial series 
\begin{equation}
 (x+1)^n=\sum\limits_{m=0}^{n}C_{m}^{n}x^{n-k}.
 \label{kineq2:binrozkl}
\end{equation}
as well as the direct ${\cal L}\{f\}=g$ and the inverse ${\cal L}^{-1}\{g\}=f$ Laplace transforms are used. Some properties of the transform are:
\begin{equation}
 {\cal L}\left\{\int_{0}^{t}f(\tau)d\tau\right\}=\frac{1}{s}{\cal L}\left\{f(t)\right\},
 \label{kineq2:Laplint}
\end{equation}
\begin{equation}
 {\cal L}^{-1}\left\{g_1(s)g_2(s)\right\}=\int_{0}^{t}f_1(t')f_2(t-t')dt'.
 \label{kineq2:Laplint5}
\end{equation}
\begin{equation}
 {\cal L}^{-1}\{ag_1(s)+bg_2(s)\}=a{\cal L}^{-1}\{g_1(s)\}+b{\cal L}^{-1}\{g_2(s)\}.
 \label{kineq2:Laplint2}
\end{equation}
\begin{equation}
 {\cal L}^{-1}\left\{g(s-c)\right\}=e^{ct}{\cal L}^{-1}\left\{g(s)\right\},
 \label{kineq2:Laplint3}
\end{equation}
\begin{equation}
\begin{array}{cc}\displaystyle
 {\cal L}^{-1}\left\{
 s^{n/2-1/2}\exp\left[-\nu^{1/2}s^{1/2}\right]
 \right\}=
 \\ \\=\displaystyle
 \frac{\exp\left[-\nu/(4t)\right]}{2^{n/2}t^{n/2+1/2}\sqrt{\pi}}\ 
 \mathrm{He}_n\left(\frac{\nu^{1/2}}{2^{1/2}t^{1/2}}\right),
\end{array}
\label{kineq2:Laplint4}
\end{equation}
where $\mathrm{He}_n(x)$ is the generalized Hermite polinomial, $n$ is integer, $\nu$ is real positive number. 
The last relation is presented on p.246 in \citet{Bateman-1954}. The first values are: $\mathrm{He}_0(x)=1$; $\mathrm{He}_1(x)=x$. Hermite polinomial and the generalized Hermite polinomial are related as
\begin{equation}
 \mathrm{H}_n\left(\frac{x}{\sqrt{2}}\right)=
 \left(\sqrt{2}\right)^n \mathrm{He}_n(x).
\end{equation}

There is a decomposition (Weisstein E. {\it MathWorld -- A Wolfram Web Resource} at http://mathworld.wolfram.com/HermitePolynomial.html) 
\begin{equation} 
 \mathrm{H}_{n}(x+y)=(\mathrm{H}+2y)^{n},
 \label{kineq2:Herm3}
\end{equation}
where $\mathrm{H}^{n}\equiv \mathrm{H}_n(x)$.
It follows from here that 
\begin{equation} 
 \mathrm{H}_{n}(x)=({\cal H}+2x)^{n},
\end{equation}
where ${\cal H}^{n}\equiv {\cal H}_n$, or
\begin{equation} 
 \mathrm{H}_{n}(x)=
 \sum_{m=0}^{n}C_{m}^{n}{\cal H}_m (2x)^{n-m},
 \label{kineq2:Herm2}
\end{equation}
with ${\cal H}_n=\mathrm{H}_{n}(0)$ to be Hermite numbers.
It can be shown that
\begin{equation} 
\begin{array}{cc}\displaystyle
 \sum_{m=0}^{n}C_{m}^{n}y^m\mathrm{H}_{m+1}(x)=
 \\ \\=\displaystyle
 y^n\mathrm{H}_{n+1}[x+1/(2y)]-y^{n-1}\mathrm{H}_{n}[x+1/(2y)].
\end{array}
\label{kineq2:HermCmA}
\end{equation}
In order to prove this, one has to use (\ref{kineq2:Herm3}) and property
\begin{equation} 
 \mathrm{H}_{n+1}(x)=\left(2x-d/dx\right)\mathrm{H}_{n}(x).
\end{equation}

\bsp
\label{lastpage}
\end{document}